\documentclass{article}
\usepackage{fullpage}
\usepackage{graphics,psfrag,epsfig}
\usepackage{amsfonts,latexsym,eucal,amsmath,amsthm,amssymb}

\begin{document}

\newcommand{\rum}{\rule{0.5pt}{0pt}}
\newcommand{\rub}{\rule{1pt}{0pt}}
\newcommand{\rim}{\rule{0.3pt}{0pt}}
\newcommand{\numtimes}{\mbox{\raisebox{1.5pt}{${\scriptscriptstyle \times}$}}}
\newcommand{\optprog}[2]
{%
  \noindent\mbox{}\\[0cm]
  \noindent\fbox{%
  \begin{minipage}{0.955\linewidth}
    \mbox{}\\[-0.5cm]
    #1\\[#2]
  \end{minipage}
  }
  \noindent\mbox{}\\[-0.2cm]
}

\renewcommand{\refname}{References}

\twocolumn[%
\begin{center}
{\Large\bf The Computational Complexity of the Traveling Salesman Problem \rule{0pt}{13pt}}\par
\bigskip
Craig Alan Feinstein \\ {\small\it 2712 Willow Glen Drive, Baltimore, Maryland 21209\rule{0pt}{13pt}}\\
\raisebox{-1pt}{\footnotesize E-mail: cafeinst@msn.com, BS"D}\par
\bigskip\smallskip
{\small\parbox{11cm}{%
\bigskip \noindent \textbf{Abstract:} In this note, we show that the Traveling Salesman Problem cannot be
solved in polynomial-time on a classical computer.

\bigskip \noindent \textbf{Disclaimer:} This article was authored
by Craig Alan Feinstein in his private capacity. No official support or endorsement by the U.S. Government
is intended or should be inferred.\rule[0pt]{0pt}{0pt}}}\bigskip
\end{center}]{%

\noindent Consider the following well-known $\mathsf{NP}$-hard problem:

\bigskip\noindent \textbf{Traveling Salesman Problem} - A traveling salesman starts at city $1$,
travels to cities $2,\dots,n-1$ in any order that the salesman chooses, and then ends his trip in city $n$.
Let us denote $\delta(i,j)$ to be the distance from city $i$ to city $j$. The goal of the Traveling
Salesman Problem is to find the minimum total distance possible for the traveling salesman to travel.
There are no restrictions on the possible distances $\delta(i,j)$ between each of the cities other than
the requirement that each $\delta(i,j)$ is a positive integer and $\delta(i,j)=\delta(j,i)$
\cite{b:CLR90,b:PS82,b:W08}.

\bigskip We give a simple proof that no deterministic and exact algorithm can solve the Traveling Salesman
Problem in $o(2^n)$ time:

For any nonempty subset $S \subseteq \{2,\dots,n\}$ and for any city $i \in S$, let us define $\Delta(S,i)$
to be the length of the shortest path that starts at city $1$, visits all cities in the set $S-\{i\}$, and 
finally stops at city $i$. Then the Traveling Salesman Problem is equivalent to the problem of computing
$\Delta(\{2,\dots,n\},n)$. Clearly, $\Delta(\{i\},i)=\delta(1,i)$ and
$$
\Delta(S,i)\,= \min \{ \Delta(S-\{i\},j) + \delta(j,i) \mid j \in S-\{i\} \},
$$
\noindent when $|S| \geq 2$.

This recursive formula cannot be simplified, so the fastest way to compute $\Delta(\{2,\dots,n\},n)$ 
is to apply this recursive formula to $\Delta(\{2,\dots,n\},n)$. Since this 
involves computing $\Delta(S,i)$ for all $\Theta(2^n)$ nonempty subsets $S \subseteq \{2,\dots,n-1\}$ and each 
$i \in S$, we obtain a lower bound of $\Theta(2^n)$ for the worst-case running-time of any deterministic and 
exact algorithm that solves the Traveling Salesman Problem.

This lower bound is confirmed by the fact that the fastest known deterministic and exact algorithm which solves 
the Traveling Salesman Problem was first published in 1962 and has a running-time of $\Theta^*(2^n)$ \cite{b:HK62,b:W08}.

}

\end{document}